# When concept-based XAI is imprecise: Do people distinguish between generalisations and misrepresentations?

Romy Müller

*Faculty of Psychology, Chair of Engineering Psychology and Applied Cognitive Research, TUD Dresden University of Technology, Dresden, Germany*

Corresponding author:

Romy Müller

Chair of Engineering Psychology and Applied Cognitive Research

TUD Dresden University of Technology

Helmholtzstraße 10, 01069 Dresden, Germany

Email: romy.mueller@tu-dresden.de

Phone: +49 351 46335330

ORCID: 0000-0003-4750-7952

## Abstract

Concept-based explainable artificial intelligence (C-XAI) can let people see which representations an AI model has learned. This is particularly important when high-level semantic information (e.g., actions and relations) is used to make decisions about abstract categories (e.g., danger). In such tasks, AI models need to generalise beyond situation-specific details, and this ability can be reflected in C-XAI outputs that randomise over irrelevant features. However, it is unclear whether people appreciate such generalisation and can distinguish it from other, less desirable forms of imprecision in C-XAI outputs. Therefore, the present study investigated how the generality and relevance of C-XAI outputs affect people's evaluation of AI. In an experimental railway safety evaluation scenario, participants rated the performance of a simulated AI that classified traffic scenes involving people as dangerous or not. These classification decisions were explained via concepts in the form of similar image snippets. The latter differed in their match with the classified image, either regarding a highly relevant feature (i.e., people's relation to tracks) or a less relevant feature (i.e., people's action). Contrary to the hypotheses, concepts that generalised over less relevant features were rated lower than concepts that matched the classified image precisely. Moreover, their ratings were no better than those for systematic misrepresentations of the less relevant feature. Conversely, participants were highly sensitive to imprecisions in relevant features. These findings cast doubts on the assumption that people can easily infer from C-XAI outputs whether AI models have gained a deeper understanding of complex situations.

# 1. Introduction

Adequate decision-making in safety-critical systems requires human and artificial agents to understand the meaning of situations. For instance, when driving a train, it is necessary to rapidly recognise danger. Such danger can manifest in perceptually distinct events involving different objects types (e.g., children playing soccer next to railway tracks vs. lorry blocking a level crossing). Despite these differences, dangerous situations share an important characteristic: It is not the object per se that causes danger but its behaviour in a particular context. Thus, if AI models are to support or even replace human decision-makers, it is insufficient for them to merely classify concrete objects (e.g., child, lorry) based on their low-level visual features. Instead, they need to be able to semantically classify scene images into abstract categories (e.g., dangerous)[1] based on what is happening in the situation (e.g., actions, relation to tracks). The technologies enabling such high-level scene understanding are still in their infancy, but research in this area is rapidly gaining momentum (for a review see Martinez Pandiani & Presutti, 2023).

A major challenge is to make it transparent how such systems arrive at their decisions. This is because the semantic classification of abstract categories is more ambiguous than that of concrete objects. Abstract categories can be highly subjective and culturally dependent (Martinez Pandiani, 2024). For instance, there is large interindividual variation in whether and why expert train drivers evaluate traffic scenes as dangerous (Müller & Schmidt, 2025). Obviously, such subjectivity impedes a clear and consistent labelling of image datasets. However, if datasets are systematically biased or if other factors keep AI models from working as intended, these models could make their decisions for the wrong reasons (e.g., inferring danger from the mere presence of a person). Therefore, explaining AI classifications of abstract categories will be paramount as task complexity increases. A promising approach is concept-based explainable artificial intelligence (C-XAI). These methods can reveal what an AI model has seen in an image, for instance by showing sets of similar snippets from other images (henceforth called concepts). These concepts are automatically selected because they evoke activation patterns in an AI's neurons similar to those evoked by particular areas in the classified image. This similarity suggests that the AI sees something in the image that resembles the concepts it provides as an explanation.

However, it is unclear how humans interpret such concepts and what inferences they draw from them. One important issue is the role of imprecision in the match between classified image and concepts. In the present study, imprecision denotes that some features of the classified image are not reflected in the concept images. Such imprecision is a doble-edged sword as it might either suggest that the AI is not working correctly or that it is able to generalise – a capability that is highly desirable in the classification of abstract categories. In the present study, such "benign generalisations" are defined as concepts that randomise over irrelevant features, thereby showing that these features do not matter. But given that people encounter benign generalisations, do they recognise and appreciate them? This entails three subquestions. First, do they evaluate benign generalisations as positively (or perhaps even more positively) as concepts that match the classified image precisely? Second, can they distinguish them from harmful generalisations that randomise over relevant features? And third, can they distinguish them from other types of benign (in the sense of non-critical) imprecisions such as systematic misrepresentations of irrelevant features?

[1] In the Cognitive Science literature, they are usually called "abstract concepts". However, to avoid confusion, the present article refers to them as "categories" and instead reserves the term "concept" for the outputs of concept-based XAI methods.

These questions were addressed in the exemplar domain of railway safety. A simulated AI classified images as dangerous or not, and its decisions were explained via C-XAI. Participants had to rate the performance of the AI, verbally explain their ratings, and label the concepts. Taken together, the present study makes the following contributions:

- It investigates how the generality and relevance of C-XAI concepts affect people's evaluation of AI models. Specifically, it assesses how these evaluations differ between concepts that either match the classified image precisely or generalise over relevant and irrelevant features.
- Qualitative data from verbal reports elucidate how people interpret the concepts and which aspects of these concepts drive their evaluations of AI performance.
- More generally, the study examines how people use C-XAI outputs that explain abstract categories (e.g., danger) via high-level features of visual scenes (e.g., relations between objects). In this way, it extends contemporary C-XAI research which has predominantly focused on extracting singular object properties but not relations between objects (i.e., unary rather than non-unary predicates). Understanding how people use more complex concepts paves the way for explaining relations.

Before describing the study in detail, a brief overview of C-XAI will be provided, focusing on the special requirements that arise when semantically classifying abstract categories. In this context, the problems and merits of imprecision will be specified. Based on literature from Cognitive Science, it will be discussed how people might deal with such imprecisions. Specifically, the challenges implied in the need to detect general principles in sets of specific images will be highlighted.

# 2. Theoretical background

## 2.1 Concept-based XAI for image classification and scene understanding

Explaining AI decisions via C-XAI is among the most promising approaches to opening the black box of deep neural networks (for reviews see Lee et al., 2024; Poeta et al., 2023; Shams Khoozani et al., 2024). In the context of image classification, C-XAI methods go beyond traditional saliency maps that merely show which image areas have contributed to an AI decision. Instead, they can also show what the AI has seen in these areas, and thus help uncover the representations learned by these complex models (Achtibat et al., 2023; Bau et al., 2017; Dreyer et al., 2025; Fel et al., 2023).

However, first and foremost, explanations need to make sense to human users. How easily can people interpret concept-based explanations and how accurately can they draw inferences from them? To answer these questions, user studies apply different strategies (for an overview see Poeta et al., 2023). Some studies assess understanding by asking people to verbally label the concepts (e.g., Wang et al., 2023; Zhang et al., 2021; Zhou et al., 2015). Some studies gather subjective evaluations of concept quality criteria such as meaningfulness, completeness, coherence, factuality, or utility (e.g., Dreyer et al., 2025). Finally, some studies use more objective measures of human performance such as the rate of correct responses in a particular task. For instance, they assess whether concepts are helpful in identifying the features an AI model has used or in detecting the biases it is prone to (Achtibat et al., 2023; Adebayo et al., 2022; Colin et al., 2022; Qi et al., 2021).

Contemporary C-XAI methods typically extract individual object properties (i.e., unary predicates). Thus, C-XAI research in image classification is predominantly concerned with explaining concrete categories based on low-level visual features like colour, shape, or location. For instance, when classifying birds, one concept might consist of image snippets showing rounded beaks, while another one might show red chests. Obviously, this narrow focus is also reflected in user studies. Accordingly, our knowledge about human understanding and interpretation of concepts is limited to these concrete cases. However, in complex decision-making tasks that rely on understanding the abstract meaning of situations, it is often not just concrete features that matter. For instance, categorising railway scenes as dangerous requires much more than mere object identification (Müller & Schmidt, 2025). Instead, AI models need to pick up what is happening, based on high-level features like actions and relations, in order to "see the intangible" (Martinez Pandiani & Presutti, 2023). Accordingly, C-XAI methods should provide evidence for (or against) such high-level scene understanding. To this end, they need to extract more complex features like relations between objects (i.e., non-unary predicates) to provide information about actions in context. However, currently we do not know how people interpret such complex concepts and how this affects their evaluation of AI models.

One important issue in this regard is how precisely the concepts shown by a C-XAI method match the classified image. Consider a scene with children playing soccer next to railway tracks, and an AI model classifying this as dangerous. A precise match would be if concepts showed children playing soccer next to railway tracks. However, the match could also be less precise. Some imprecisions eliminate key features responsible for the abstract category and thus change the meaning of the situation. For instance, they might not retain the relation to railway tracks (e.g., children playing soccer in different areas). Conversely, other imprecisions might actually be beneficial, namely when the concepts bring out the essence of a situation more clearly by means of generalisation. They can do so by retaining relevant features while randomising over less relevant ones, such as the specific action (e.g., children engaged in various activities next to railway tracks). As famously stated by Jorge Luis Borges (1962, p.115): "To think is to forget a difference, to generalize, to abstract." If people realised that an AI model is able to look beyond image-specific details, this might give them the impression that it has gained a deeper understanding of the situation, and thus evaluate it more favourably. Note that this hypothesis is at odds with the common aim to maximise the coherence of C-XAI concepts (i.e., all snippets showing the same thing) and to interpret high coherence ratings in user studies as an indicator of C-XAI quality (Poeta et al., 2023; Shams Khoozani et al., 2024). Some variation in C-XAI concepts may actually be beneficial, especially in high-level semantic classification, because it shows that the AI is able to ignore irrelevant features. But will people recognise and appreciate such generalisations? This will critically depend on whether they can pick up subtle differences between concepts. To generate predictions that can be tested empirically, it is necessary to understand how humans generalise concepts.

## 2.2 Human generalisation of actions and relations

To distinguish between helpful and harmful imprecisions in C-XAI outputs, people would need to recognise *generalisations* in concept images. This cannot be taken for granted, neither for individual images nor for sets of multiple images. For individual images, a reason to doubt whether people will recognise general concepts in them is that images encourage a low level of construal (Rim et al., 2015). That is, when people see images of objects, they are more likely to categorise them in a narrow, specific manner than when they see the same objects presented as words. However, images of simple objects differ from images of complex scenes. When people describe and label scene images, the terms they use are fairly general and interpretive (Greisdorf & O'Connor, 2002; Jörgensen, 1998; Laine-Hernandez & Westman, 2006; Rorissa & Iyer, 2008). In the context of C-XAI, an additional challenge arises as each concept consists of multiple images. Thus, even benign generalisations might not be obvious, because people need to actively generate them by integrating over a set of diverse image snippets. However, this might be feasible, because in fact, people describe groups of images more abstractly than individual ones (Rorissa, 2008). That said, it is unclear whether findings from image description studies are transferrable to the context of C-XAI. First, all these studies used fairly diverse images, which might have encouraged participants to generalise. Second, when participants had to label groups of images, they generated these groups by themselves (Greisdorf & O'Connor, 2002; Rorissa, 2008). Most likely, they generated groups for which they can easily find a common theme. Third, these group labels were abstract in many ways (e.g., sports, landscapes, sad) but seldom referred to actions.

This raises the question whether people will spontaneously recognise generalisations over *actions* depicted in C-XAI concepts. For instance, will they notice that a concept represents "being on tracks" when the individual snippets show various actions performed on the tracks? There is reason to assume that they will. First, natural images of human actions activate mental representations of abstract categories (McRae et al., 2018). For instance, when seeing an image of two children eating a corncob together, this activates the abstract action category of "sharing". Thus, people should be able to interpret C-XAI concepts in terms of their abstract meaning. Additional support stems from *Action Identification Theory*, which specifies how people represent actions on various levels of abstraction (Vallacher & Wegner, 1987). For instance, one and the same behaviour can be described as "standing on one leg on the edge of a railway track", "balancing on tracks", "being on tracks", or "putting your life in danger". A core assumption of Action Identification Theory is that people usually prefer more abstract descriptions that promote a holistic understanding of a situation's meaning. Accordingly, they might appreciate it when C-XAI concepts vary in low-level action details, as long as the meaning stays intact. However, generalising over actions is not the same as being unaware of these actions or misrepresenting them. Therefore, it needs to be tested whether people distinguish between generalisations and systematically incorrect actions depicted in C-XAI concepts.

Other generalisations are not justified. In the railway context, this presumably is the case for generalisations over a potential obstacle's *relations* to tracks. Will people notice and reject such harmful generalisations? This seems quite likely, as human thinking is genuinely relational, and many categories are determined by the relations between objects or features (Erickson et al., 2005; Gentner & Kurtz, 2005). When viewing images, people are sensitive to semantic and spatial relations between objects, which also manifests in their eye movement patterns (Boettcher et al., 2018; Hwang et al., 2011; Torralba et al., 2006). Perhaps most impressively, the relational nature of people's visual representations is showcased by the finding that their brain activity in response to natural images is well-aligned with LLM embeddings of scene captions (Doerig et al., 2022). Thus, one might expect that people are aware of the relations reflected in C-XAI concepts. On the other hand, being aware of relations is not trivial as people rarely mention relations or locations when labelling or describing images (Greisdorf & O'Connor, 2002; Jörgensen, 1998; Laine-Hernandez & Westman, 2006). However, this is likely to depend on context. In the domain of railway safety, relations are essential. Accordingly, when evaluating an AI system for person detection, the spatial relations of missed people to railway tracks were the strongest influence on participants' ratings (Müller, 2025). Walking *towards* tracks instead of *away from* them can easily turn a harmless situation into a disaster. Thus, it seems likely that people will not tolerate C-XAI concepts that imprecisely reflect a person's relation to tracks. Any potential benefits of generalisation are not expected to apply to relations.

## 2.3 Present study

How are human evaluations of AI models affected by imprecision in C-XAI concepts? The present study tested the assumption that some imprecisions are considered beneficial, while others are not. This was expected to depend on the generality and relevance of concept features. Participants had to evaluate a simulated AI that classified still images of people in the vicinity of railway tracks as dangerous or not. The AI's decisions were explained by showing which concepts it has seen in the classified image. Participants had to perform four tasks: indicate whether the AI has decided correctly, rate its performance, explain this rating, and provide a verbal label for the concept. Five concept types varied the match between the classified image and the concept's image snippets. Some concepts were *precise matches* in that they closely resembled the classified image. That is, each snippet showed people performing the same action (presumably a less relevant feature) with the same relation to tracks (presumably a highly relevant feature). Moreover, there were three types of imprecise concepts. For *benign generalisations*, the snippets varied in the less relevant actions but retained the more relevant relations to tracks. To control for the effects of systematically misrepresenting the less relevant feature rather than generalising over it, *benign confusions* showed the same incorrect action in each snippet. For *harmful generalisations*, the action always matched the classified image but the relation to tracks varied across snippets. Finally, *irrelevant feature* concepts showed snippets that only matched the classified image in a surface feature such as people's clothing. This served as a lower baseline to put the results for imprecise concepts into perspective.

Overall, it was hypothesised that people distinguish between different types of imprecise concepts when evaluating AI performance. The highest ratings were expected for both precise matches and benign generalisations, with no decrement in ratings for the latter compared to the former (H1). A finding like this would suggest that participants notice and appreciate it when an AI seems to grasp the deeper meaning of a situation. Moreover, ratings for benign generalisations were expected to be higher than those for the two other types of imprecise concepts. First, they were expected to be higher than for benign confusions (H2). This would suggest that participants notice generalisations as such, instead of simply being unaware of the depicted actions or not caring whether the AI misrepresents them. Second, ratings for benign generalisations were expected to be higher than for harmful generalisations (H3). This would indicate that participants reject an AI that seems oblivious to the essential features. Finally, the lowest ratings were expected for irrelevant feature concepts (H4), suggesting that participants reject an AI that only attends to surface features but is blind to both actions and relations. It was hypothesised that benign generalisations would be rated higher than this lower baseline, while harmful generalisations would not.

All these hypotheses presuppose that participants interpret the concepts in terms of the depicted actions and relations to tracks. To investigate what they actually saw in the concepts and how this influenced their evaluations of the AI, a qualitative content analysis was performed on the verbal concept labels and explanations of ratings.

# 3. Methods

## 3.1 Data availability

All stimuli, instructions, behavioural data, verbal explanations, and syntax files are made available via the Open Science Framework: https://osf.io/cbzrj/

## 3.2 Participants

Thirty-three participants (27 female, 6 male) took part in the study. They were recruited via the TUD Dresden University of Technology's participant pool. Participants' age ranged from 20 to 68 years (*M* = 29.2, *SD* = 12.5). All participants had normal or corrected-to-normal vision and were fluent in reading and writing German. They received partial course credit or a monetary compensation of 10 € per hour. Participants provided written informed consent and all procedures followed the principles of the Declaration of Helsinki.

## 3.3 Apparatus and stimuli

### *3.3.1 Technical setup*

The experiments ran on one of three desktop computers (screen size 24"). A standard computer mouse was used as an input device. A separate laptop was positioned next to participants for entering concept labels and explanations of their ratings in a Microsoft Excel sheet. The experiment was programmed using the Experiment Builder (SR Research, Ontario, Canada, Version 2.4.193). No actual AI was used in the study and all stimuli were generated manually. This was done to enable a fine-grained manipulation of the match between concept images and classified image, allowing for a systematic, factorial variation of AI decisions and concept types.

### *3.3.2 Instruction video*

Participants watched an instruction video that lasted 7:05 minutes and was based on a Microsoft PowerPoint presentation. It explained that the aim of the study was to investigate how people evaluate AI systems that classify railway traffic scenes as dangerous or not. Participants were informed that C-XAI was able to convey why the AI has made a particular decision. This was illustrated using the simple example of classifying an image as a dog. On consecutive slides, concepts were presented that matched the classified image more or less precisely and were more or less relevant, explaining decisions that were correct or incorrect. It was made explicit that participants' numerical ratings of AI performance may incorporate different aspects (e.g., correctness of decision, match between concepts and classified image, match between concepts and decision) and that participants were free to choose how to integrate them. Subsequently, the procedure was explained with screenshots from the experiment.

### *3.3.3 Images*

The classified images and concept images were assembled from five rail-specific datasets: RailGoerl24 (Tagiew et al., 2025), RailSem19 (Zendel et al., 2019), OSDaR23 (Tagiew et al., 2023), RAWPED (Toprak et al., 2020), and a Rail Dataset with images of track workers (Rail, 2022). All images depicted people in the vicinity of railway tracks at various distances and while performing various actions.

*Classified images*. Sixty images (plus two for practice) had ostensibly been classified by an AI (for an example see Figure 1A). They reflected six scenarios: walking towards tracks, walking on tracks, looking away close to tracks, walking away from tracks, standing on platform, and sitting on bench on platform. For each scenario, ten similar images were selected. For instance, if the scenario was "walking towards tracks", all images depicted a person walking towards tracks but the images varied with regard to surface features such as the person's identity, clothing, or movement phase. For each classified image, a highlighted version was created by overlaying the relevant people with a heatmap from red to blue. This highlight ostensibly demarked the image areas the AI had used to classify the image.

*Concept images*. Concepts ostensibly reflected what the AI had seen in the highlighted area. For each of the 60 classified images, a unique concept image consisting of five individual images was created. These were snippets taken out of railway images that were not part of the classified image set. A snippet zoomed into the image in such a way that only the relevant parts were retained (usually a person and the tracks). There were five concept types that varied in their match with the classified image (see Figure 1 for examples, Table 1 for principles of matching, and Table 2 for specific contents).

**Figure 1.** Example for classified image and five types of concept images. (A) Classified image. (B) Precise match. (C) Benign generalisation. (D) Benign confusion. (E) Harmful generalisation. (F) Irrelevant feature.

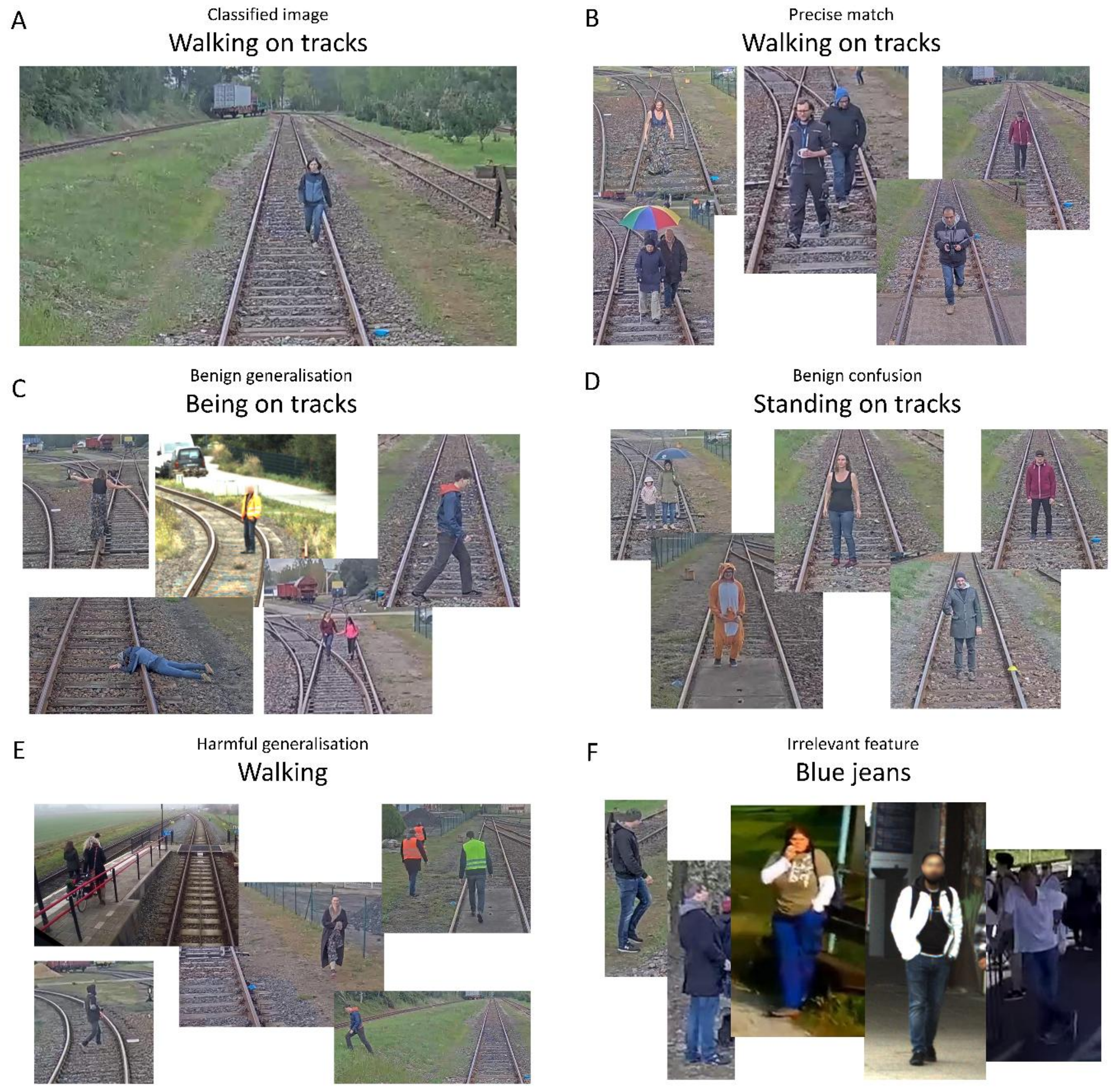

First, for *precise matches*, all five snippets depicted the same theme as the classified image (e.g., walking on tracks). That is, people's relation to the tracks as well as their actions were retained. Second, *benign generalisations* retained people's relation to the tracks but the five snippets randomised over a presumably less relevant feature. The latter could be the person's specific action (e.g., walking vs. standing on tracks), the exact cause of being unable to see the train (e.g., facing in the opposite direction vs. sight barrier), the present versus immediate future (e.g., being on tracks now vs. after two more steps), or the physical context (e.g., standing on a platform vs. far from the tracks elsewhere). Third, *benign confusions* were similar to benign generalisations in that they retained the most relevant feature (i.e., relation to tracks) but not a less relevant feature (i.e., the specific action). However, the same incorrect action was shown in each of the five snippets. For instance, when the classified image showed a person walking on the tracks, all snippets showed people standing on the tracks. Thus, benign confusions reflected systematic mistakes rather than generalisations. Fourth, for *harmful generalisations*, people's relation to the tracks varied between the five snippets, while the action was the same as in the classified image. For instance, all people were walking, but not only on the tracks but also next to them, across them, or away from them. Thus, harmful generalisations presumably were inacceptable as they randomised over an essential feature. Finally, *irrelevant feature* concepts consisted of snippets that matched the classified image in a surface feature such as the person's clothing, but neither depicted a relation to the tracks nor any systematic action.

**Table 1.** Characterisation of the match between the five concept types and the classified image with regard to the relation to tracks (relevant feature) and performed action (less relevant feature).

| | Precise match | Benign generalisation | Benign confusion | Harmful generalisation | Irrelevant feature |
|---|---|---|---|---|---|
| Relation to tracks | same | same | same | randomised | none |
| Action | same | randomised | different | same | randomised |

**Table 2.** Image themes of the classified images and concept types for all six scenarios. For irrelevant feature concepts, separate features were shown in the trial in which the AI classified the image as dangerous and not dangerous, respectively.

| Classified image | Precise match | Benign generalisation | Benign confusion | Harmful generalisation | Irrelevant feature |
|---|---|---|---|---|---|
| Walking towards tracks | Walking towards tracks | Being or going to be on tracks | Standing close to tracks | Walking | Red clothing/ Dress |
| Walking on tracks | Walking on tracks | Being on tracks | Standing on tracks | Walking | Blue jeans/ Dark coat |
| Looking away close to tracks | Looking away close to tracks | Unable to see close to tracks | Looking at train close to tracks | Looking away | Umbrella/ Yellow or orange |
| Walking away from tracks | Walking away from tracks | Being away from tracks | Standing close to tracks | Walking | Hood/ Blue jacket |
| Standing on platform | Standing on platform | Being away from tracks | Walking on platform | Standing | Sunglasses/ Pink clothing |
| Sitting on bench on platform | Sitting on bench on platform | Being on platform | Standing on platform | Sitting | Bagpack/ Basecap |

### *3.3.4 Experimental screens*

Stimuli were shown with a resolution of 1920 × 1080 px. The screens contained images, interaction elements, and white, green or red text on a black background. All text was presented in German. Three consecutive screens were shown in each trial (see Figure 2): a decision screen, a rating screen, and a prompt to provide verbal concept labels and explanations.

The *decision screen* displayed the image number, the AI decision (i.e., "dangerous" in green or "not dangerous" in red font), the classified image (1280 × 720 px), a prompt to evaluate the AI decision ("Has the AI decided correctly?"), and two buttons (a light green one on the left-hand side labelled "correct decision" and an orange one on the right-hand side labelled "incorrect decision"). The *rating screen* again provided the image number and AI decision. Additionally, it presented a highlighted version of the classified image on the left-hand side and a concept image on the right-hand side (both 960 × 540 px). The highlighted image was titled “These are the areas the AI has attended to” and the concept image was titled “Concept: This is what the AI has seen”. Below these images, a text asked “How well has the AI done its job?” A slider with the poles “very poorly” and “very well” was provided and participants could click any position to submit an integer point score between 0 and 100. In the screen prompting participants to *provide verbal concept labels and explanations*, the lower part of the rating screen (i.e., question and slider) was replaced by two questions to be answered in an Excel sheet: “What has the AI seen there (concept)?” and “Why have you provided this point score?” Moreover, there was a button to confirm that all information has been entered.

**Figure 2.** Screens shown in the experiment. (A) Decision screen. (B) Rating screen. The screen with the prompt to provide verbal concept labels and explanations is not shown as it was similar to the rating screen (see text for details).

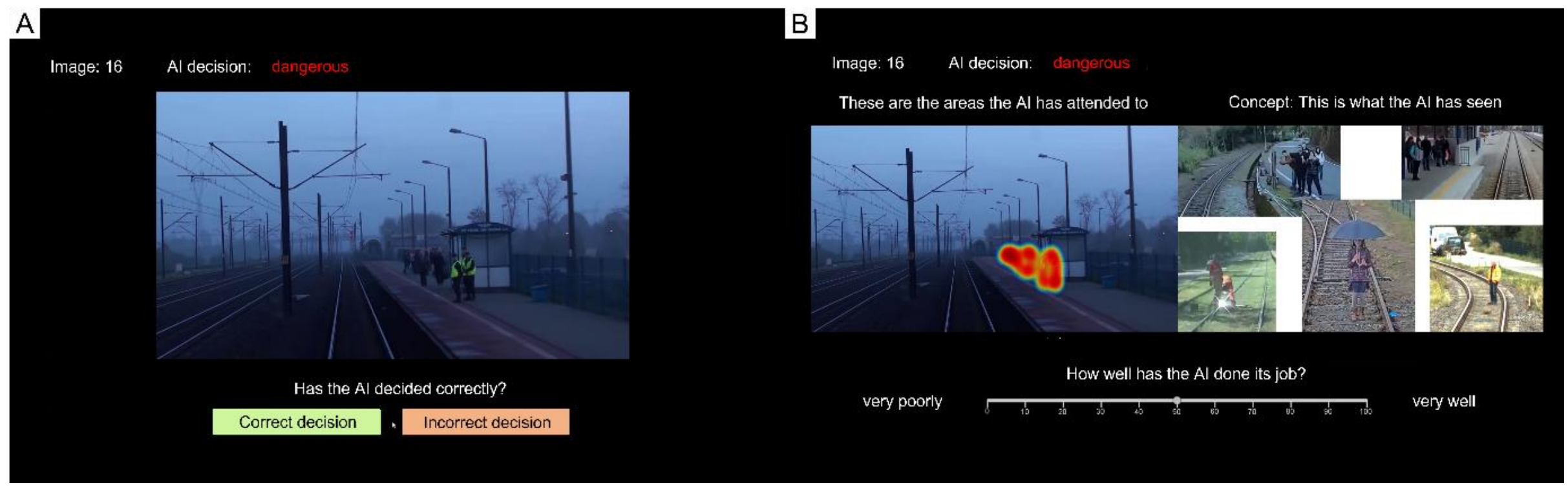

## 3.4 Procedure

Participants first provided written informed consent and demographic information and then watched the instruction video. Afterwards, they completed two practice trials and 60 experimental trials, presented in random order, with each trial showing a unique classified image and concept. The 60 trials corresponded to all possible combinations of the six scenarios, five concept types, and two AI decision outcomes (see Figure 3).

**Figure 3.** Overview of trial types. Each scenario was paired with each trial type and AI decision, resulting in 60 unique combinations that were presented in 60 trials.

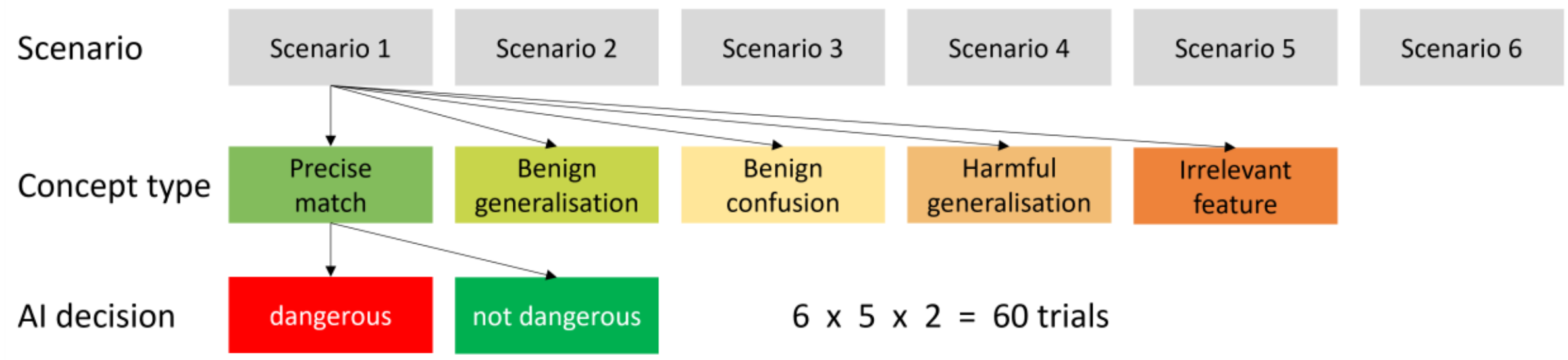

Thus, all factors were varied within participants. After starting a trial by pressing a button, participants entered the decision screen on which they indicated whether they agreed or disagreed with the AI decision. Pressing the respective button transferred them to the rating screen. They could switch back and forth between a highlighted and non-highlighted version of the classified image by pressing “X” on their keyboard, and inspect the concept images for as long as they wanted. In order to answer the question “How well has the AI done its job?”, they had to click a position on the slider. This automatically brought them to the screen prompting them to provide verbal concept labels and explanations. In a Microsoft Excel sheet, they assigned a label to the concept (consisting of a word or short phrase) and explained why they had selected their respective point score. Altogether, the experiment took between one and two-and-a-half hours.

## 3.5 Data analysis

For the first seven participants, two of the 60 trials were excluded from all analyses as they showed an incorrect concept image. For all remaining trials, three dependent variables were analysed: participants' ratings of AI performance, the labels they assigned to the concepts, and the explanations they provided for their ratings.

### *3.5.1 Ratings of AI performance*

It was analysed how participants' mean ratings of AI performance depended on the AI decision and concept type. Furthermore, the accuracy of AI decisions was included as a factor in the analysis. This is because when an AI errs in a safety-critical domain (e.g., classifies a person walking on the tracks as not dangerous), this is likely to have detrimental effects on participants' evaluations, possibly eliminating any effects of concept type. Thus, the data were split according to decision accuracy as perceived by participants (i.e., agreement with the AI) rather than "true" accuracy. This is because participants were expected to differ in their assessments of danger. Some might even consider a person walking on the tracks as not dangerous, for instance if they assumed that the person had seen the train and would therefore leave in time. Accordingly, one and the same AI decision might be perceived as correct by some participants and incorrect by others. In consequence of splitting the data, two participants had a missing value in one of the cells, because they never disagreed with the AI in a particular condition (regardless of whether the AI decided that one and the same scenario type was dangerous or not). These two values were replaced by the mean of all other participants for the respective cell. To statistically analyse participants' ratings, a 2 (*agreement with AI: agree, disagree*) × 2 (*AI decision: dangerous, not dangerous*) × 5 (*concept type: precise match, benign generalisation, benign confusion, harmful generalisation, irrelevant feature*) repeated-measures ANOVA was conducted. Only main effects and interactions involving the factor concept type will be reported. If sphericity was violated, a Greenhouse-Geisser correction was applied and the degrees of freedom were adjusted accordingly. To determine statistical significance, an alpha level of $p < .05$ was used and all pairwise comparisons were performed with Bonferroni correction.

### *3.5.2 Concept labels*

It was analysed how participants verbally labelled the concepts used by the AI. Two participants were excluded as they labelled the classified images instead of the concepts. For the remaining participants, their concept labels were coded according to whether they contained information about relations, actions, and surface features. For instance, the label "person on tracks" only contains a relation, while the label "person walking on tracks" contains an action and a relation. For each information content, the percentage of trials was calculated in which the respective information was included. Concept labels were analysed using a 5 (*concept type: precise match, benign generalisation, benign confusion, harmful generalisation, irrelevant feature*) × 3 (*information content: relations, actions, surface features*) repeated-measures ANOVA.

*3.5.3 Explanations of ratings*

Presumably, other aspects of concepts besides the ones manipulated in the present study contribute to participants' evaluation of AI performance. To determine what these aspects are, participants' explanations of their ratings were analysed. As the focus of the present study was on concepts, statements pertaining to the AI decision were only coded as to whether the decision was explicitly mentioned as a reason or not. For statements about concepts, it was coded whether the explanation contained any information about the concept at all, whether it explicitly evaluated the concept's quality, and what aspects of the concept it referred to. In the quantitative analysis, statements about concept quality were not included, because participants' verbal strategies differed between concepts that they thought were good versus bad or mixed. For good concepts, participants often merely provided an explicit overall evaluation (e.g., the concept is good, the concept fits perfectly) without specifying why. Conversely, for bad and mixed concepts, they often did not make this overall evaluation explicit but instead elaborated on the problematic aspects (e.g., the concept does not consider whether a person is on the tracks or next to them). Accordingly, simply counting the instances of explicitly evaluating a concept would lead to a distorted picture, with many more good evaluations than bad or mixed ones. Therefore, only the analysis of specific information contents is reported. The categories used for coding these information contents were developed inductively while coding participants' explanations. Overall, the following types of categories emerged: abstract features of the concepts, features that were manipulated experimentally, other features of people, features of the physical context, and other features that do not fit into any of the previous categories. The detailed contents of these categories will be described in Section 4.3. For all categories, the percentage of trials in which they were mentioned was analysed descriptively.

# 4. Results

## 4.1 Ratings of AI performance

By comparing participants' ratings of AI performance between concept types, it was assessed whether they recognise and appreciate generalisation in C-XAI outputs. To this end, participants' mean ratings of AI performance were averaged across the six scenarios and analysed depending on concept type and its interactions with AI decision and agreement with the AI. There was a main effect of concept type, $F(2.5,80.3) = 91.089$, $p < .001$, $\eta_p^2 = .740$, an interaction of concept type and agreement with the AI, $F(3.1,175.8) = 16.439$, $p < .001$, $\eta_p^2 = .339$, as well as an interaction of concept type and AI decision, $F(3.1,99.2) = 3.647$, $p = .014$, $\eta_p^2 = .102$. Moreover, the three-way interaction was significant, $F(4,128) = 9.640$, $p < .001$, $\eta_p^2 = .232$ (see Figure 4 and Table 3).

For the factor concept type, a basic pattern emerged that will be relevant in most of the interactions described below: All concept types differed from each other, all *p*s < .001, except for benign generalisations and benign confusions, $p > .9$. This pattern can be specified as follows. Ratings were highest when concepts matched the classified image precisely (62.8) and lowest when they showed irrelevant features (31.9). When concepts were imprecise, ratings were similar for benign generalisations and benign confusions (54.8 and 56.6), but considerably lower for harmful generalisations (39.9). Thus, participants did not seem to care whether the AI generalised over actions or systematically misrepresented them, as long as the relation to tracks was not violated.

The interaction of concept type and agreement with the AI indicated that the pattern described above occurred when participants agreed with the AI decision. In this case, the ratings for all concept types differed from each other, all *p*s < .006, except for benign generalisations and benign confusions, *p* > .9. When participants disagreed with the AI, ratings no longer differed between precise matches, benign generalisations, and benign confusions, all *p*s > .164, while they were lower for harmful generalisations and still lower for irrelevant feature concepts, all *p*s < .046.

The interaction of concept type and AI decision indicated that the pattern described for concept type as a whole was present when the AI decided that a situation was dangerous. In this case, all concept types differed from each other, all *p*s < .005, except for benign generalisations and benign confusions, *p* > .9. When the AI decided that the situation was not dangerous, the pattern looked similar, but now benign confusions were not only on par with benign generalisations but even with precise matches, both *p*s > .9. Moreover, harmful generalisations were on par with irrelevant feature concepts in terms of receiving low ratings, *p* = .759. All other comparisons were significant, all *p*s < .001.

Finally, the three-way interaction indicated the following. First, when participants agreed with the AI that a situation was dangerous, the basic pattern was present again: All concept types differed from each other, all *p*s < .005, except for benign generalisations and benign confusions, *p* > .9. Similarly, when participants agreed with the AI that a situation was not dangerous, most concept types differed from each other, all *p*s < .001, except for benign generalisations and benign confusions, *p* > .9, as well as harmful generalisations and irrelevant feature concepts, *p* > .9. Third, when participants disagreed with the AI deciding that a situation was dangerous (i.e., participants thought it was not dangerous), they provided similar ratings for precise matches, benign generalisations, and benign confusions, all *p*s > .9, and similar ratings for harmful generalisations and irrelevant feature concepts, *p* > .9, while harmful generalisations and irrelevant feature concepts yielded lower ratings than the other three concept types, all *p*s < .002. Fourth, when participants disagreed with the AI deciding that a situation was not dangerous (i.e., participants thought it was dangerous), no differences between concept types were significant, all *p*s > .052, except that the ratings for irrelevant feature concepts were lower than those for all other concept types, all *p*s < .021.

**Figure 4.** Ratings of AI performance depending on agreement with the AI, AI decision, and concept type. Ratings are presented for trials in which participants (A) agreed or (B) disagreed with the AI. Error bars represent standard errors of the mean.

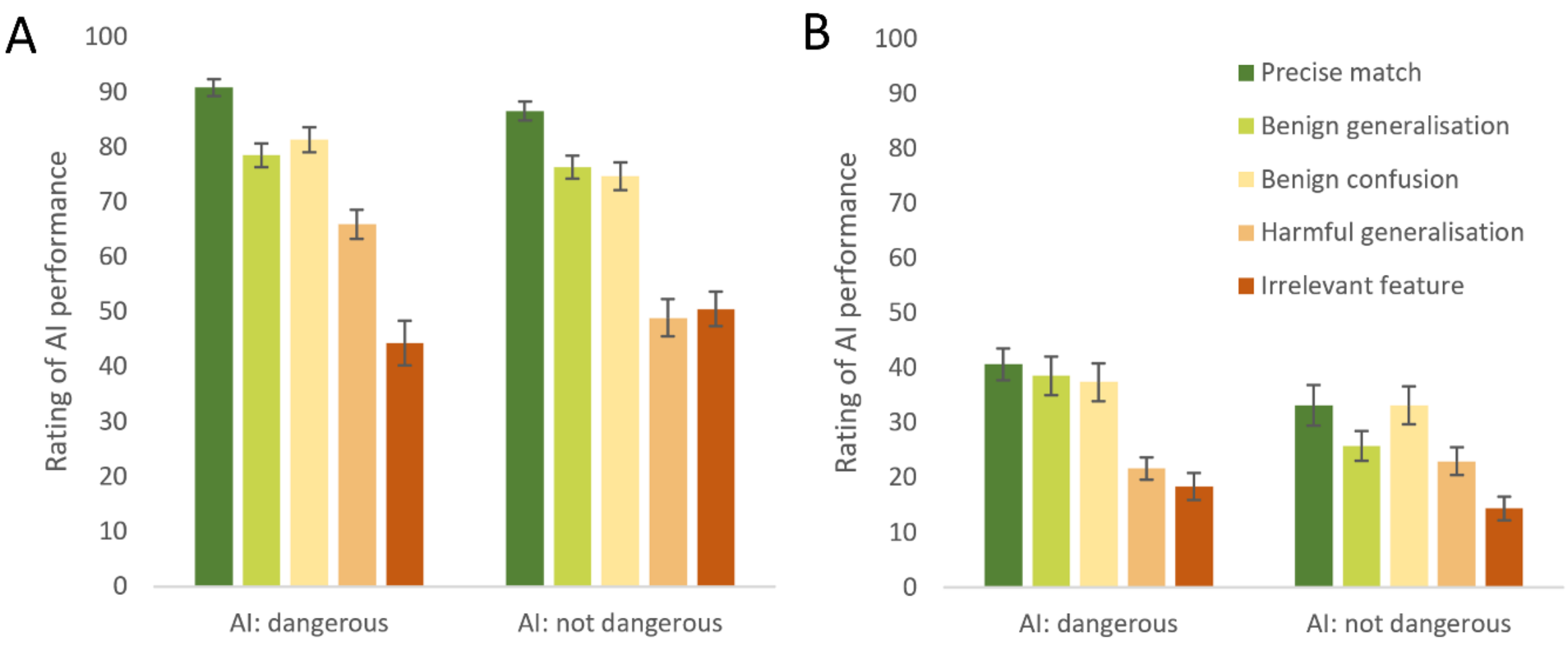

**Table 3.** Mean values and standard deviations (in parentheses) for ratings of AI performance (as point scores from 0-100) and information contained in participants' concept labels (percentage of labels containing the respective information).

| | | Precise match | Benign generalisation | Benign confusion | Harmful generalisation | Irrelevant feature |
|---|---|---|---|---|---|---|
| Ratings of AI performance | | | | | | |
| Agree with AI | AI: dangerous | 90.8 (8.9) | 78.5 (12.5) | 81.3 (13.1) | 65.9 (15.0) | 44.3 (23.5) |
| | AI: not dangerous | 86.5 (10.0) | 76.3 (12.3) | 74.7 (14.5) | 48.9 (19.6) | 50.5 (18.0) |
| Disagree with AI | AI: dangerous | 40.7 (16.6) | 38.5 (20.3) | 37.4 (19.8) | 21.7 (11.7) | 18.3 (14.2) |
| | AI: not dangerous | 33.1 (21.2) | 25.7 (15.9) | 33.1 (20.3) | 23.0 (14.5) | 14.4 (12.5) |
| Information content of concept labels (%) | | | | | | |
| Relations | | 93.8 (10.5) | 93.8 (10.1) | 96.4 (8.3) | 87.1 (20.5) | 12.1 (17.9) |
| Actions | | 55.6 (27.5) | 36.8 (26.8) | 35.1 (28.6) | 37.1 (25.0) | 30.9 (30.8) |
| Surface features | | 0.3 (1.5) | 0.0 (0.0) | 0.6 (2.5) | 0.3 (1.5) | 41.9 (34.5) |

## 4.2 Concept labels

By analysing how participants verbally labelled the concepts, it was assessed what they actually see in the sets of image snippets. For the five concept types, the percentage of participants' concept labels was determined that contained information about relations, actions, and surface features. There was a main effect of information content, $F(1.5,45.4) = 156.373$, $p < .001$, $\eta_p^2 = .839$, a main effect of concept type, $F(2.4,71.8) = 41.654$, $p < .001$, $\eta_p^2 = .581$, as well as an interaction of both factors, $F(2.7,82.0) = 93.193$, $p < .001$, $\eta_p^2 = .756$ (see Figure 5A and Table 3).

The main effect of information content indicated that relations were included in the majority of labels (76.7 %), while actions were mentioned less often (39.1 %) and surface features were rare (8.6 %), all *p*s < .001. The main effect of concept type indicated that for precise matches, labels contained more information than for all other concept types, for irrelevant feature concepts they contained less, all *p*s < .001, and for the three types of imprecise concepts, the information content did not differ between them, all *p*s > .9. However, the interaction indicated that the differences between concept types varied with the type of information. Relations were mentioned similarly often for precise matches, benign generalisations, benign confusions, and harmful generalisations, all *p*s > .9, but less often for irrelevant feature concepts than for all other concept types, all *p*s < .001. Actions were mentioned more often for precise matches than for all other concept types, all *p*s < .002, while the latter did not differ from each other, all *p*s > .9. Finally, surface features were mentioned most often for irrelevant feature concepts, all *p*s < .001, but only in less than 1 % of the trials and thus similarly rarely for all other concept types, all *p*s > .09.

A look at the interindividual differences in participants' concept labels revealed several interesting observations (see Figure 5B). First, participants consistently included relations to tracks for all concept types except irrelevant feature concepts. Second, they largely varied in whether they included actions. Some almost never did (e.g., they only assigned labels like "person on tracks"), while others included actions more often, but to varying degrees. Third, surface features were omitted almost entirely by almost all participants, except for irrelevant feature concepts. Although the latter were defined by these surface features, participants varied in the degree to which they included them (i.e., some simply assigned labels like “person” or “man”, without further specification).

**Figure 5.** Information included in participants’ concept labels. (A) Mean percentage of relations, actions, and surface features depending on concept type. Error bars represent standard errors of the mean. (B) Interindividual differences. Each line represents one participant.

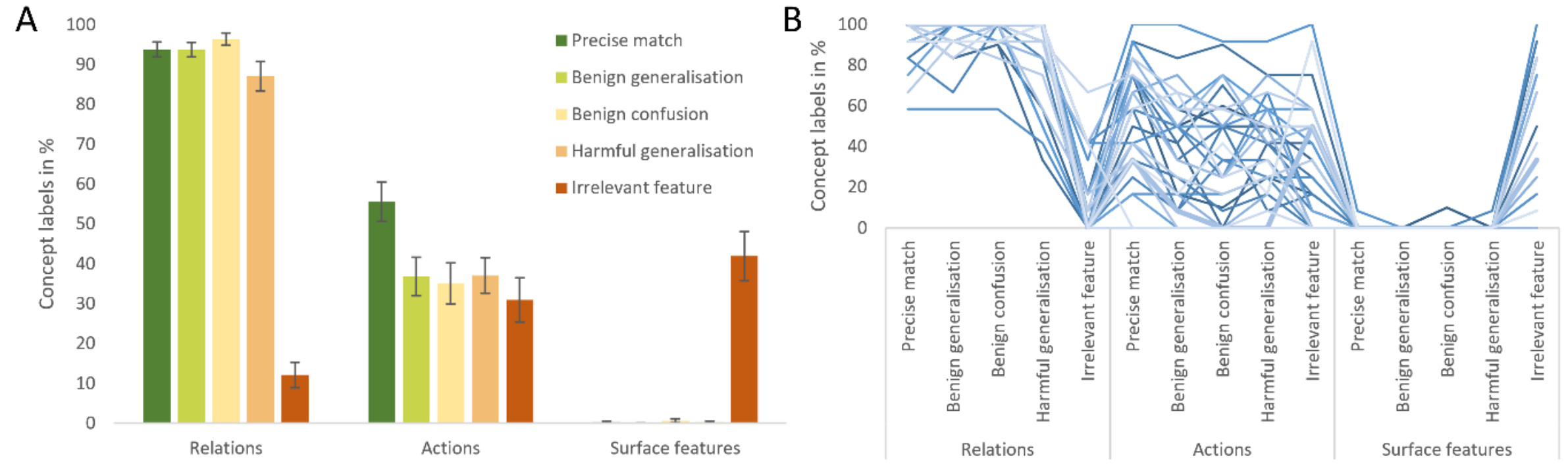

## 4.3 Explanations of ratings

By analysing how participants explained their ratings of AI performance, it was assessed what features of the concept images they considered important. When explaining their ratings, participants referred to the AI decision in 74.4 % and to the concept in 87.7 % of the trials. When their explanations mentioned aspects of the concept, five types of features were distinguished (see Table 4): features describing abstract aspects of the concept images (i.e., imprecision, heterogeneity, danger), features that were experimentally manipulated in the present study (i.e., relation, action, irrelevance), other features of people (i.e., identity, number, gaze & vision, mental states, workwear), features of the physical context (i.e., infrastructure, no tracks, environment), and other features that were only mentioned too rarely to deserve their own category (i.e., only by an individual participant or less than three times in total). The results are presented in Figure 6.

**Table 4.** Concept-related reasons that participants gave when explaining their ratings of AI performance. Percentages indicate in how many explanations the respective aspect was mentioned.

| Category | Explanation contents | Examples | Percent |
|---|---|---|---|
| | | Abstract features | |
| Imprecision | How closely the concept matches the classified image | Concept images are too imprecise, concept does not fit exactly | 7.2 |
| Heterogeneity | How similar the concept’s image snippets are to each other | Concept images show very different things, images are inconsistent | 4.2 |
| Danger | How dangerous the situations depicted in the concept are | Concepts are more dangerous than the image, all concepts are from non-dangerous situations | 6.5 |
| | | Manipulated features | |
| Relation | Relation to tracks (including movement direction) | Parallel to tracks was recognised correctly, concept did not attend to proximity to tracks | 12.9 |
| Action | Type of action (excluding movement direction) | Some people are standing and some are walking, concepts attend more to sitting than to waiting on platform | 4.5 |
| Irrelevance | Surface features and explicit judgments of irrelevance for other features | AI has attended to red clothing, whether people are standing or walking is less relevant | 4.0 |
| Other person features | | | |
| Identity | Whether the mere presence of people is used, what type of people are used | AI has only attended to the person, concept ignores that it is a child | 3.6 |
| Number | How many people are present | Concept does not consider the number of people, concept does not fit to danger for single person | 2.1 |
| Gaze & vision | Where people are facing and whether they can see the train | Not all people are turned towards the train, person does not look at the train | 1.0 |
| Mental states | Assumptions about people's attention, intentions, and other mental states | People in concept look as if they knew what they are doing, AI does not attend to intention to cross the tracks | 0.4 |
| Workwear | Whether people are wearing safety vests (indicating that they are workers) | Some people are wearing safety vests, concepts do not reliably recognise safety vest | 0.5 |
| Physical context features | | | |
| Infrastructure | Relevant infrastructure that is included or missing | Concepts without level crossing, concept intermixes platform and bare tracks | 5.4 |
| No tracks | Absence of tracks | There are no tracks in the concepts, concepts do not focus on danger due to tracks | 2.4 |
| Environment | Weather, time of day, underspecified references to context | One concept image with snow does not fit, concept does not sufficiently attend to context | 1.0 |
| Other features | | | |
| Various | Legality, time, image quality | Concepts could consider whether the situation is legal, snippets are too small | 0.6 |

Abstract features of concepts were mentioned in a considerable share of participants' explanations. Participants commented on the concepts' imprecision (usually criticising that concepts were not sufficiently precise but never that they were too precise), on their heterogeneity (mostly evaluating it as too high), and on whether the concepts reflected dangerous situations (often assessing whether they were more or less dangerous than the classified image). Besides these abstract features, participants also attended to several aspects of the concepts' content. First, for the features manipulated in the present study, the results mirrored those from the previous analyses. That is, relations to tracks were the most important aspect by far, while actions were mentioned less often. At first glance, it seems surprising that actions were only mentioned about as much as irrelevance, although each concept necessarily depicted actions but only 20 % focused on surface features. Presumably, irrelevance was used comparably often because it did not only count references to surface features but also assertions that another feature was irrelevant (e.g., people's specific actions). Furthermore, participants referred to features of people other than the manipulated ones. Most of these statements referred to people's presence and identity (e.g., that concepts only focused on the mere presence of a person, describing the type of person). Moreover, participants commented on the number of people present in the concepts versus classified images, and a few mentioned the direction of people's gaze, their inferred mental states, or whether they wore safety vests (classifying them as workers who presumably know what to do). Complementary to features of people, a considerable share of explanations mentioned the physical context, most often referring to the infrastructure (e.g., criticising that not all concept snippets showed platform scenes). Participants also usually mentioned the absence of tracks for irrelevant concepts, and sometimes mentioned other environmental factors (e.g., darkness, snow). Other features were rarely mentioned, in most cases only by an individual participant (e.g., complaining about the technical quality of concept images). Taken together, these explanations show that the concepts features manipulated in the present study reflect aspects that are considered important by participants, but that several additional features contributed to their evaluations of AI performance.

**Figure 6.** Percentage of trials in which participants' explanations of their ratings mentioned particular features of the concept images. Abstract features are presented in yellow, manipulated features in dark purple, other person features in light purple, physical context features in blue, and other features in grey. Error bars represent standard errors of the mean.

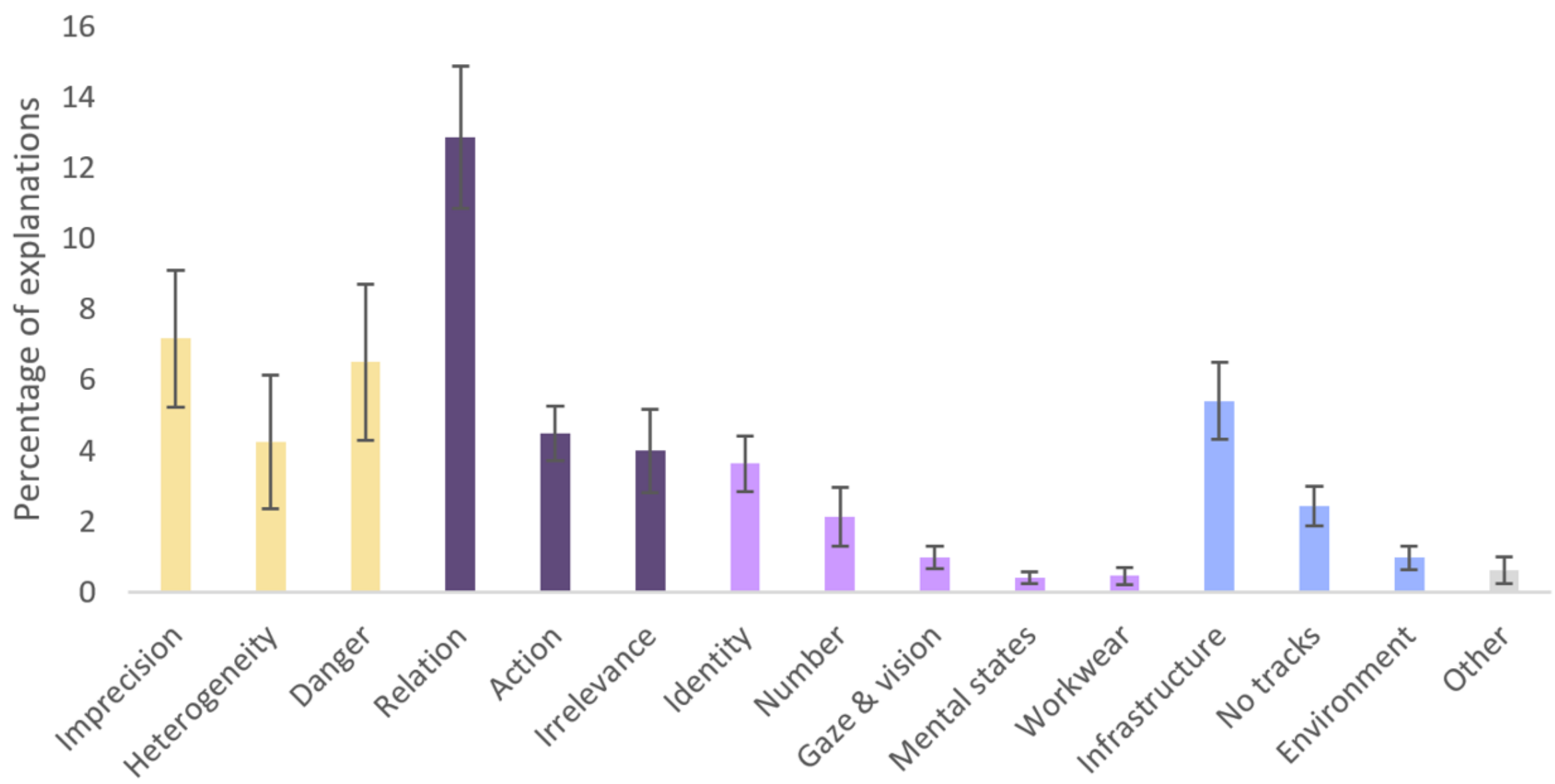

## 5. Discussion

Do people recognise and appreciate generalisation in the concepts presented by C-XAI methods? This was investigated for the context of classifying abstract categories in the exemplar domain of railway safety evaluation. Participants rated a simulated AI that classified traffic scene images with people in the vicinity of the tracks as dangerous or not. These decisions were explained via C-XAI concepts that varied in their match with the classified image, either in a less relevant feature like the specific action, or in the highly relevant relation to the tracks. Contrary to the hypotheses, participants did not appreciate concepts that generalised beyond the details of a classified image by randomising over less relevant features. For these benign generalisations, AI performance was rated lower than for precise matches (contradicting H1) and similar to benign confusions that systematically depicted a wrong action (contradicting H2). As expected, benign generalisations led to higher ratings than harmful generalisations that ignored the relation to tracks (supporting H3) and higher ratings than concepts that focused on irrelevant surface features (supporting H4). Taken together, these findings suggest that people do notice the presence and relevance of imprecisions in C-XAI concepts, and readily incorporate this into their evaluations of AI models. However, it seems like they either are unaware of the representational causes of imprecisions, or indifferent to them. That is, as long as it is not an essential feature that is represented imprecisely, it does not matter whether the AI's representations are generalised or systematically mistaken. The following sections will discuss why no generalisation benefits were found in the present study and what features of concepts participants considered. Moreover, limitations of the study will be made transparent and future directions will be highlighted.

## 5.1 No benefits of generalisation

Perhaps the most remarkable finding of the present study is that participants did not distinguish between benign generalisations and benign confusions, as long as these imprecisions concerned a less relevant feature. At least three explanations might account for this absence of a generalisation benefit. First, it might stem from a *lack of awareness* of the less relevant feature. Participants might simply not have noticed what exactly the people depicted in the concept images were doing and whether this varied across snippets. A finding in favour of this speculation is that less than 40 % of participants' concept labels mentioned actions, in line with previous findings that people rarely mention actions when labelling groups of images (Greisdorf & O'Connor, 2002; Rorissa, 2008). However, another finding seems incompatible with a lack of awareness: Participants often mentioned the heterogeneity of concepts when explaining their ratings of AI performance. Interestingly, they almost exclusively did so in a critical way, remarking that the concepts were not sufficiently coherent. This directly leads to the second possible explanation.

Participants' disapproval of benign generalisations might have been a matter of *task representation*. The fact that participants were so critical of heterogeneity suggests that the best concepts, from their perspective, were those with the least variation. Their underlying idea might have been that if an AI is designed to select similar images, this similarity should be maximal. This resonates with an assumption underlying many user studies on C-XAI, namely that a high coherence within concepts is generally desirable (Poeta et al., 2023). If variation was generally perceived as negative, this might partly explain why benign generalisations were rated no higher than benign confusions, which at least were coherent. Participants might not have understood that some imprecisions can be beneficial. This might change when adjusting the instruction to transfer more knowledge about AI. Participants could be made aware of the fact that overfitting is a serious problem of AI systems, that robustness is highly desirable, and that the ability to generalise is something worth striving for (Muttenthaler et al., 2024).

According to a third explanation, the lack of generalisation benefits might be a matter of *degree*. Strictly speaking, as soon as concepts are not identical to the classified image, they can be considered generalisations. Even precise matches generalised over several features, both of people (e.g., identity, clothing, exact movement phase) and the physical context (e.g., area, time of day). Participants clearly paid attention to these features, as indicated by the verbal explanations of their AI ratings. Perhaps from participants' perspective, there is a sweet spot of optimal generalisation, and the present study's precise matches happened to hit it. This would imply that concepts can also be too precise in retaining features of the classified image. Just like objects, actions have a preferred basic level of description (e.g., walking rather than tiptoeing) (Győri, 2019) which people readily use to describe images (Rorissa & Iyer, 2008). Generally speaking, people prefer more abstract levels of action identification (Vallacher & Wegner, 1987). Accordingly, they might not appreciate concepts that match the actions too closely (e.g., all snippets showing a person with the left foot lifted by 45° during a step to the right-hand side of the image). Any potential aversion to such overprecision could not have been detected in the present study, because even the actions shown in precisely matching concepts were more diverse. Thus, future studies could gradually vary the level of action abstraction in C-XAI concepts to investigate whether there is indeed an optimal amount of generalisation.

## 5.2 What features of concepts matter to people?

A consistent finding reflected in all analyses was the powerful role of *relations to tracks*. First, when they were amiss, participants' ratings of AI performance went down. Especially when the AI classified the situation as not dangerous (and participants agreed!), ratings dropped to the lower baseline of completely irrelevant concepts. Second, relations were by far the most prominent concept feature mentioned in participants' verbal explanations of their ratings. Third, they were featured in almost all of participants' concept labels. This is not trivial, because people rarely use relations or locations to describe or label images (Greisdorf & O'Connor, 2002; Jörgensen, 1998; Laine-Hernandez & Westman, 2006). Particularly, when people need to be concise, this almost entirely eliminates locations from their descriptions (Laine-Hernandez & Westman, 2006). Thus, it is remarkable that relations to tracks were so prominent here and that participants were highly sensitive to an AI inaccurately representing them (cf. Müller, 2025).

This raises the question to what extent the dominance of relations over actions is context-specific. In the exemplar domain of railway safety, danger is tightly linked to a potential obstacle's relation to the tracks. Presumably, the conclusion that generalising over actions is acceptable, while generalising over relations is not, would not hold in all other contexts. For instance, in many social settings, it is essential to consider the specific action despite identical spatial relations (e.g., wrestling vs. hugging). Details of actions also are paramount in many other domains such as sports or criminology. Thus, transferring the present findings to other domains might only be possible on a more abstract level. An abstract conclusion would be that people are highly sensitive to imprecisions in relevant features, but neither appreciate generalisations of less relevant features nor care when these features are represented incorrectly. Future studies could test this hypothesis in contexts where actions are the most relevant feature. This could be tested using large image datasets like MS COCO, for which 140 common visual actions are annotated for 10.000 images (Ronchi & Perona, 2015).

Besides the actions and relations manipulated in the present study, participants considered *other features of concepts,* as revealed by their verbal explanations. For instance, they put considerable focus on a person's identity. Some even punished the AI when the concepts consisted of different types of people, for instance when the classified image showed a man and the snippets included women, children, or workers. The same was true for the physical context. Participants often complained that some snippets showed platforms while the classified image did not, or vice versa. Some of these features, and many others that varied within concepts, can certainly affect danger. For train drivers it is highly relevant indeed whether a person is a worker or child, or whether this person is located on a platform (Müller & Schmidt, 2025). To represent this diversity of relevant features, some C-XAI methods show several sets of snippets (i.e., several concepts) at the same time and make the relevance of each set explicit (Achtibat et al., 2023; Fel et al., 2023). Future studies should investigate whether the presentation of multiple concepts affects people's perspectives on generalisation.

The aspects of concepts that matter to people may also *vary between individuals*. In the present study, this was reflected in participants' concept labels. While they almost unanimously included relations on a regular basis, they considerably varied in how often they included actions. Similar variability has been reported in other studies. Most closely related to the present findings, when people described motion events, they consistently included the motion's path but varied immensely in whether they included its manner (Montero-Melis, 2021). Moreover, when people assigned labels to groups of images, this resulted in a multitude of unique descriptions with little overlap (Greisdorf & O'Connor, 2002; Rorissa, 2008). Thus, it is not uncommon that people differ in how they perceive and describe concepts. Future studies might scrutinise how this depends on stable interindividual differences. One candidate is expertise. The present study was conducted with novice participants, who assigned little relevance to people's actions and mental states. This is not what train drivers do, who habitually attend to people's actions, body language, and inferred mental states to evaluate whether situations are dangerous (Müller & Schmidt, 2025). Thus, the relevance of concept features might change with professional experience. Another candidate is language. There is considerable cross-linguistic variation in how people describe images (Berger & Ponti, 2024). While this has different cultural and linguistic reasons, even the mere grammatical structure of a language can influence how people describe motion events. For instance, speakers of languages in which it is optional to include the manner of motion are less likely to verbally distinguish between fine-grained manner categories (Malt et al., 2014). Similarly, speakers of languages that do not have a grammatical indicator of ongoingness tend to report fewer actions but more goals or endpoints (Athanasopoulos & Bylund, 2013; von Stutterheim & Nüse, 2003). In some studies, the influence of language went even deeper and affected which image areas people attended to, how they categorised events, or what events they remembered (Athanasopoulos & Bylund, 2013; Gennari et al., 2002; Kersten et al., 2010; von Stutterheim et al., 2012). Thus, one might hypothesise that language can also affect how people evaluate generalisation in C-XAI concepts.

### 5.3 Limitations and future directions

Several limitations of the present study constrain the inferences to be drawn from it. A first set of limitations concerns the balance between internal and external validity. On the one hand, the study was not as strictly controlled as it could have been. For instance, benign generalisations were multi-dimensional, with some concepts generalising over actions and others over causes of an ability, brief timescales, or physical contexts. Future research could vary concept types more stringently to systematically tease apart the effects of feature type, feature relevance, similarity to the classified image, coherence between image snippets, and several other potential influences.

On the other hand, for the sake of experimental control, the study was limited in its external validity. All stimuli were generated manually, and no actual AI or C-XAI algorithms were applied. This was necessary to implement the experimental variations of interest. However, the study's prototypical imprecisions in C-XAI concepts would probably not be produced in the same way by a real C-XAI algorithm. Thus, future studies should investigate how people deal with realistic imprecisions in the outputs of actual C-XAI methods. As these methods do not yet exist for the classification of abstract categories like danger and non-unary predicates like relations between objects, initial studies might have to start in the context of object classification.

Another set of limitations concerns the study's design decisions, some of which were arbitrary but nonetheless may have influenced the results. One is the implementation of the verbal concept labelling task. Participants had to provide brief labels consisting of a word or short phrase. It is known that the specific description task changes the contents of image descriptions (Jörgensen, 1998; Laine-Hernandez & Westman, 2006). Thus, it can be speculated that allowing for longer verbal descriptions would have led participants to include more information about actions.

A final limitation is that the problem identified in the present study is closely linked to a particular type of C-XAI: using example images from the dataset to show what an AI model has learned. Such examples necessarily depict specific situations, which can then be generalised via randomisation over less relevant features. In the future, it should be investigated how people deal with other XAI approaches such as hypericons that are generated from the features a deep neural network has used to classify abstract categories (Martinez Pandiani et al., 2023). Moreover, it is also possible to go beyond purely visual concepts. A promising future perspective is to assess the role of language in dealing with imprecise concepts. Some C-XAI methods provide verbal information in their concepts (e.g., Kim et al., 2018). For instance, adding the phrase “being on tracks” might change people’s reactions to generalisations. This is because language in general and labels in particular can serve as a means of abstraction (Lupyan & Lewis, 2019; Rim et al., 2015) and highlight commonalities between diverse exemplars that might otherwise go unnoticed (Davis & Yee, 2019). Thus, concept labels should make generalisations more easily identifiable and distinguishable from systematic misrepresentations. In this way, future studies could zoom in on the question of how people deal with generalisations when it is clear that they are aware of them in the first place.

### 5.4 Conclusion

The outputs of C-XAI methods do not always match the classified image precisely, and the present study asked how people deal with that. Particularly, do they recognise and appreciate generalisations? The findings suggest otherwise. Participants preferred precise matches and did not distinguish between generalisations and systematic misrepresentations of image features as long as they did not concern the most relevant features. This suggests that interpreting C-XAI is not trivial. It may not be self-evident to users that robustness has a price. Future studies should leverage knowledge about human categorisation and generalisation, both to develop C-XAI methods and to investigate how people interpret and use them. If it is true that “To think is to forget a difference, to generalize, to abstract.” (Borges, 1962, p.115), we should enable people to recognise and appreciate these abilities in AI models.

## Acknowledgments

I want to thank Judith Schmidt, Carsten Knoll, Sascha Weber, and Florian Funk for valuable discussions of the experimental setup. This work was supported by the German Centre for Rail Traffic Research (DZSF) at the Federal Railway Authority within the project “Explainable AI for Railway Safety Evaluations (XRAISE)”.